\documentclass[preprint,superscriptaddress,showpacs,preprintnumbers,amsmath,amssymb,aps,pre]{revtex4-2}
\usepackage[utf8]{inputenc} 
\usepackage{epsfig}
\usepackage{natbib}
\usepackage{booktabs}
\usepackage{amsmath}
\usepackage{times}
\usepackage{epstopdf} 
\usepackage{graphicx}
\usepackage{color}

\usepackage{afterpage}

\begin{document}

\title{Final state sensitivity in noisy chaotic scattering}

\author{Alexandre R. Nieto}
\email[]{alexandre.rodriguez@urjc.es}
\affiliation{Nonlinear Dynamics, Chaos and Complex Systems Group, Departamento de
F\'{i}sica, Universidad Rey Juan Carlos, Tulip\'{a}n s/n, 28933 M\'{o}stoles, Madrid, Spain}

\author{Jes\'{u}s M. Seoane}
\affiliation{Nonlinear Dynamics, Chaos and Complex Systems Group, Departamento de
F\'{i}sica, Universidad Rey Juan Carlos, Tulip\'{a}n s/n, 28933 M\'{o}stoles, Madrid, Spain}

\author{Miguel A.F. Sanju\'{a}n}
\affiliation{Nonlinear Dynamics, Chaos and Complex Systems Group, Departamento de
F\'{i}sica, Universidad Rey Juan Carlos, Tulip\'{a}n s/n, 28933 M\'{o}stoles, Madrid, Spain}
\affiliation{Department of Applied Informatics, Kaunas University of Technology, Studentu 50-415, Kaunas LT-51368, Lithuania}

\date{\today}

\begin{abstract}
	
	The unpredictability in chaotic scattering problems is a fundamental topic in physics that has been studied either in  purely conservative systems or in the presence of weak perturbations. In many systems noise plays an important role in the dynamical behavior and it models their internal irregularities or their coupling with the environment. In these situations the unpredictability is affected by both the chaotic dynamics and the stochastic fluctuations. In the presence of noise two trajectories with the same initial condition can evolve in different ways and converge to a different asymptotic behavior. For this reason, even the exact knowledge of the initial conditions does not necessarily lead to the predictability of the final state of the 
	system. Hence, the noise can be considered as an important source of unpredictability that cannot be fully understood using the
	conventional methods of nonlinear dynamics, such as the exit basins and the uncertainty exponent. By adopting a probabilistic point of view, we develop the concepts of probability basin, uncertainty basin and noise-sensitivity exponent, that allow us to carry out both a quantitative and qualitative analysis of the unpredictability on noisy chaotic scattering problems.
	
\end{abstract}

\pacs{05.45.Ac,05.45.Df,05.45.Pq}
\maketitle
\newpage
\section{Introduction} \label{sec:Introduction}

Unpredictability is a fundamental topic in nonlinear science, and it is a 
consequence of the sensitive dependence to initial conditions, inherent to chaotic systems. However, there are several ways to understand unpredictability, and for each of them many concepts, methods and tools have been 
developed. The unpredictability can be defined as the difficulty to predict the evolution of the trajectories. With this point of view, the expansion entropy \cite{Hunt}, the Kolmogorov-Sinai entropy \cite{Kolmogorov,Sinai}, the topological entropy \cite{Adler}, and other measures have been developed. Nevertheless, in certain physical situations such as chaotic scattering problems \cite{Seoane13}, we are interested in the asymptotic behavior rather than the evolution of the trajectories. In this case, the unpredictability is understood as the difficulty to predict the final state of a trajectory that starts with a particular initial condition. The asymptotic behavior can be a fixed point or a chaotic attractor in dissipative systems, a leak in chaotic Hamiltonian maps or an opening in the potential in open
Hamiltonian systems. The most common tools to analyze the
unpredictability in this kind of systems are the basins of attraction \cite{NusseD} and the exit basins \cite{Contopoulos02}. A basin of attraction of a dissipative system is defined as the set of initial conditions 
that are attracted to a certain attractor. On the other hand, an exit basin of a Hamiltonian system is the set of initial conditions that escape through a particular exit of the system. When two different exits (or attractors) coexist, two basins appear, separated by a smooth or a fractal boundary. The knowledge
of the structure of the basin boundary, together with other
characteristics of the basins, allows to understand and quantify the unpredictability of the system. For this purpose, the uncertainty exponent \cite{Grebogi83} and the basin entropy \cite{Daza16} are some of 
the most powerful tools when working on fundamental aspects of physical systems. On the other hand, in the case of applied physics and some fields 
of engineering, many methods to characterize the dynamical integrity \cite{Thompson89} of the basins of attraction have been developed. Some examples are the anisometric local integrity measure (ALIM) \cite{Belardinelli18}, the integrity factor (IF) \cite{Rega05}, and the local integrity measure (LIM) \cite{Soliman89}.

Recently, much work has been done on the unpredictability of chaotic scattering. In particular, current research has shifted
the focus from purely conservative systems to the effects of
relativistic corrections \cite{Bernal20}, small
perturbations as dissipation \cite{Seoane07}, and periodic forcing \cite{Blesa14,Nieto18}. However, the noise is ubiquitous in nature, so that important lines of research have arisen to elucidate the noise effects in chaotic scattering \cite{Rodrigues,Silva,Mills,Altmann,Seoane08}. Accordingly, in this manuscript we carry on a research
on noisy chaotic scattering, mainly focusing our attention on developing numerical and theoretical techniques that allow to characterize the final 
state sensitivity.

Our research arises from the observation that some of the tools and methods used to study unpredictability in chaotic scattering are useless in the presence of stochastic fluctuations. As good examples, we can mention the exit basins and the uncertainty exponent. These tools are based on the 
high sensitivity of the asymptotic behavior to initial conditions or parameters.  However, if we consider a small amount of noise the system is not deterministic anymore and the situation changes drastically. In this scenario, two exactly identical initial conditions can evolve in a different way and converge to a different asymptotic state. Hence, in presence of 
noise we are not interested anymore in the  neighborhood of the initial condition but in the initial condition itself.

Next, we develop some tools and numerical methods that
can help to understand and quantify the effects of noise on the
unpredictability of chaotic scattering systems. With this goal in mind, and without loss of generality, we have used two paradigmatic open Hamiltonian systems. Nevertheless, we expect that the main results could be general in chaotic scattering problems in which the noise models the effect of internal irregularities or the coupling of the system with the environment. We presume potential applications to several fields of physics such as celestial mechanics \cite{Bernal20,Assis} and plasma physics \cite{Mathias,Mathias2}, among others. In particular, both the exit basins and the 
noise play an important role in the magnetic behavior in tokamaks \cite{VianaS,Turri,Alkesh}. Other possible applications
include chaotic scattering problems in different fields of science, such as medicine \cite{Schelin}, biology \cite{Scheuring} or chemistry \cite{Ezra,Kawai}.

The structure of this paper is as follows. In Sec.~\ref{Model}, we describe the models of this work, the H\'{e}non-Heiles system and the four-hill 
potential, both in presence of a source of
additive uncorrelated Gaussian noise. The description of the
concepts and methods to compute the probability basin and the
uncertainty basin is carried out in Sec.~\ref{BP} and
Sec.~\ref{BU}, respectively. In the latter, we also focus our
attention on the relation between the structure of the uncertainty basin and the stable manifold of the chaotic saddle. In Sec.~\ref{QFSS}, we develop the concept of noise-sensitivity exponent as a tool to quantify the final state sensitivity to noise. Finally, in Sec.~\ref{Conclusions}, we present the main conclusions of this work.

\section{Description of the models}\label{Model}

One of the models that we have used to illustrate the ideas and methods of this manuscript is the Hénon-Heiles system~\cite{HH64}. This Hamiltonian system appeared for the first time in $1964$ as a model of a galactic potential. The Hamiltonian is characterized by a nonlinear axisymmetric potential and is given by

\begin{equation} \label{eq:HH_Hamiltonian}
	{\cal{H}}=\frac{1}{2}(\dot{x}^2+\dot{y}^2)+\frac{1}{2}(x^2+y^2)+x^2y-\frac{1}{3}y^3,
\end{equation}
where $x$ and $y$ denote the space coordinates, and $\dot{x}$ and $\dot{y}$ are the the momentum coordinates.

\begin{figure}[htp]
	\centering
	\includegraphics[width=0.8\textwidth,clip]{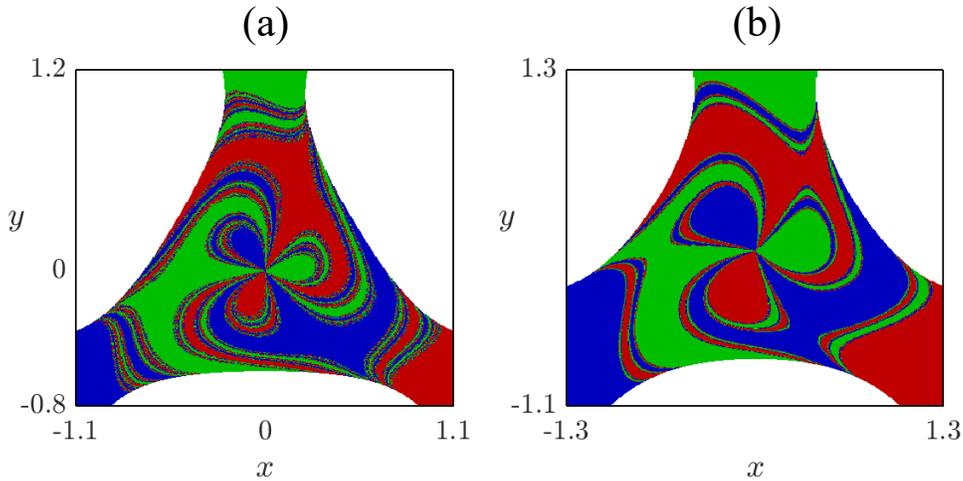}
	\caption{Exit basins of the Hénon-Heiles system in physical space with energy (a) $E=0.25$ and (b) $E=0.45$. The colors red, green and blue refer to initial conditions leading to the three exits of the potential: Exit 1 ($y\to\infty$), Exit 2 ($x,y\to-\infty$) and Exit 3 ($x\to\infty,y\to-\infty$).}
	\label{fig1}
\end{figure}

The system has been extensively studied due to the fractal
exit basins and the wide variety of dynamical behaviors existing for different values of the energy~\cite{Aguirre01,Barrio08}. To visualize the system in the context of the unpredictability analysis, in Fig.~\ref{fig1} we show two exit basins for different values of the energy $E=0.25$ and $E=0.45$. When the energy of the system is increased, the fractal basin boundaries become thinner and less fractal, with a consequent 
reduction on the unpredictability. This decreasing can be quantified by using both the uncertainty exponent and the basin entropy.

For the purposes of this research, we have included in the equations
of motion a source of additive uncorrelated Gaussian noise, so the
equations of motion become \cite{Seoane08}

\begin{equation} \label{eq:eq motion}
	\begin{aligned}
		\dot{p} & = -x - 2xy + \sqrt{2\xi}\eta_x(t) \\
		\dot{q} &= -y - x^2 + y^2 + \sqrt{2\xi}\eta_y(t),
	\end{aligned}
\end{equation}
where $\xi$ is the intensity of the noise and $\eta_x(t)$, $\eta_y(t)$ are white noise processes with variance $\sigma^2 = 2\xi$ and mean $\mu=0$ .

For comparison purposes, we have also used the four-hill system \cite{Bleher,Zotos}, whose potential consists of four hills located at $(x, y) = 
(\pm1, \pm1)$ and its Hamiltonian is given by:

\begin{equation} \label{eq:4H}
	{\cal{H}}=\frac{1}{2}(\dot{x}^2+\dot{y}^2)+x^2y^2e^{-(x^2+y^2)}.
\end{equation}

Under the same considerations that we have mentioned in the case of the Hénon-Heiles, the equations of motion in presence of noise read

\begin{equation} \label{eq:eq motion4H}
	\begin{aligned}
		\dot{p} & = 2xy^2(x^2-1)^{-(x^2+y^2)} + \sqrt{2\xi}\eta_x(t) \\
		\dot{q} &= 2x^2y(y^2-1)^{-(x^2+y^2)} + \sqrt{2\xi}\eta_y(t).
	\end{aligned}
\end{equation}

\begin{figure}[htp]
	\centering
	\includegraphics[width=0.8\textwidth,clip]{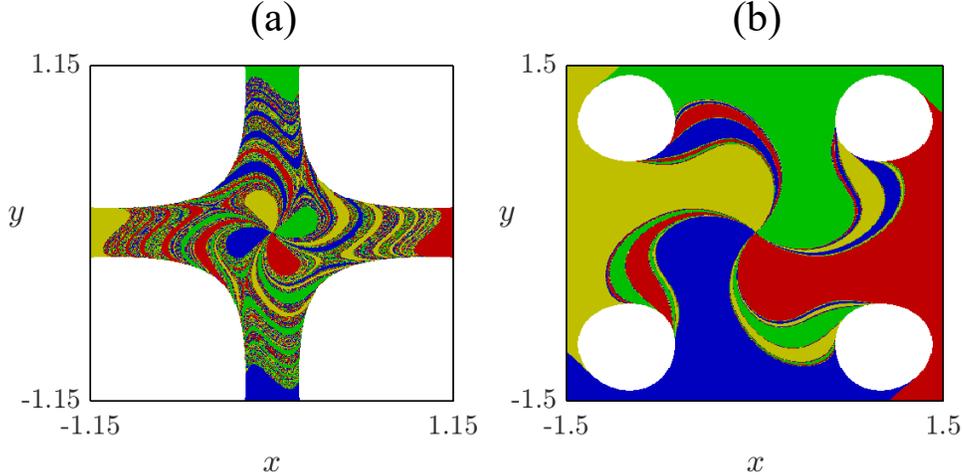}
	\caption{Exit basins of the four-hill system in the physical space with energy (a) $E=0.01$ and (b) $E=0.10$. The colors green, yellow, blue and red refer to initial conditions leading to the four exits of the potential: Exit 1 ($y\to\infty$), Exit 2 ($x\to-\infty$), Exit 3 ($y\to-\infty$) and Exit 4 ($x\to\infty$).}
	\label{fig2}
\end{figure}

To visualize the system, we represent the exit basins for different energies $E=0.01$ and $E=0.1$ in Fig.~\ref{fig2}. In the low energy case, the basin boundary occupies an important part of the phase space. While in the high energy case the basin boundary is thinner, letting space to extensive safe regions of high predictability. If we increase even more the 
energy until $E=1/e^2$, the trajectories evolve on the top of the hills 
and the scattering becomes non-chaotic.

We have solved the stochastic differential equations for the Hénon-Heiles and the four-hill system by using the stochastic Heun method \cite{Kloeden}, as was previously used in Refs.~\cite{Seoane08,Bernal13}. In both systems, the numerical methods gave stable and convergent solutions. To ensure the effectiveness of the method, we have also compared the results by using the numerical schemes of the Euler-Maruyama method and the stochastic fourth-order Runge-Kutta method.

\section{Probability basins}\label{BP}

\begin{figure}[htp]
	\centering
	\includegraphics[width=0.8\textwidth,clip]{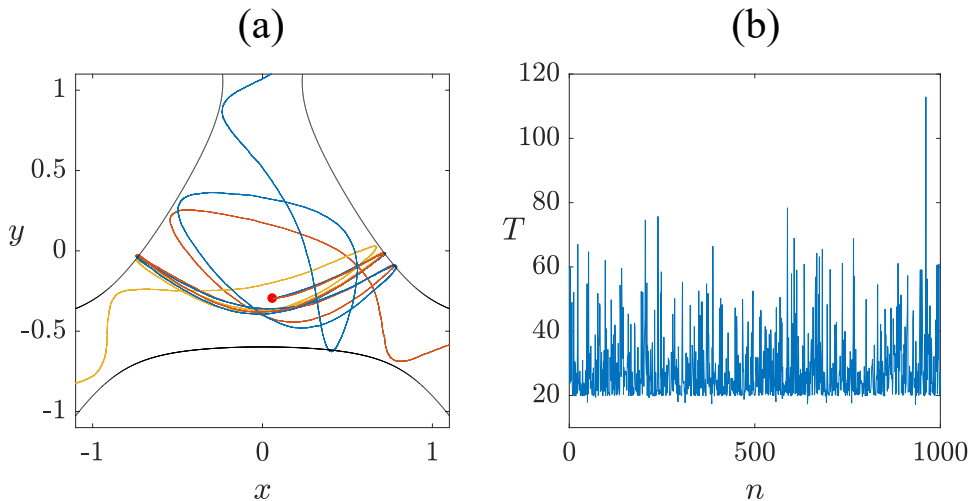}
	
	\caption{Escape dynamics of the noisy Hénon-Heiles system with $E=0.25$ and $\xi=10^{-5}$, showing (a) three trajectories in the physical space launched from the same initial conditions escaping through different exits, and (b) escape time distribution of $100$ launchings with the same initial condition.}
	\label{fig3}
\end{figure}

Perhaps the most obvious and at the same time interesting effect of noise 
on the escape dynamics is that identical initial conditions can escape through different exits and spend a different time in the scattering region. To illustrate this, we represent three trajectories with the same initial condition escaping through different exits of the potential of the Hénon-Heiles system in Fig.~\ref{fig3}(a). In addition, the escape times obtained for different launchings of the same initial condition are represented in Fig.~\ref{fig3}(b). This figure reminds us the very usual scattering function. However, here no coordinate is varied and the apparently disordered escape times are due to the effects of the noise. Even with a perfectly known initial condition, we cannot say by which exit the particle will escape and we can only establish that the trajectory will remain in the scattering region during a certain range of time $T\in[19,116]$.

In this situation, where the main characteristics of the scattering
process (asymptotic behavior and escape time) may change in
different simulations, it is necessary to adopt a probabilistic
approach in order to understand the escape dynamics. Hence, the
escape of a trajectory should be understood as a probability of
escape. Additionally, the escape times of an initial condition should be
analyzed in terms of the average and the variance of the individual
escape times. Regarding unpredictability, one of the most useful
portraits is the structure of the exit basins. However, in presence of noise this representation lacks meaning since for a given initial condition, the exit is not uniquely defined. So, in noisy systems we do not have a 
unique exit basin representation, we might have a different one for each different simulation we carry out. As it has been reported in
Refs.~\cite{Seoane08,Bernal13}, the exit basins appear smeared or blurred, and the basin boundary becomes shaded off due to the effects of noise. Hence, a deeper understanding of the system is not possible by using the exit basins. Nevertheless, we can construct a similar representation that 
we call \textit{probability basin}, in which every initial condition in the grid does not have an associated asymptotic behavior, but a probability of escaping through a particular exit. Consequently, we will have as many probability basins as there are exits in the system. For the computation of the probability basins, we start fixing a value of the noise intensity and establishing a grid of $P$ initial conditions in the phase space region that we want to analyze. Then, we compute $N$ times the trajectory 
of the very same initial condition and we detect the exit through which they escape. Once we know the exits of all $N$ trajectories, we can establish the probability of each exit. By repeating this procedure for all $P$ 
initial conditions we will obtain the probability basins.

\begin{figure}[htp]
	\centering
	\includegraphics[width=0.8\textwidth,clip]{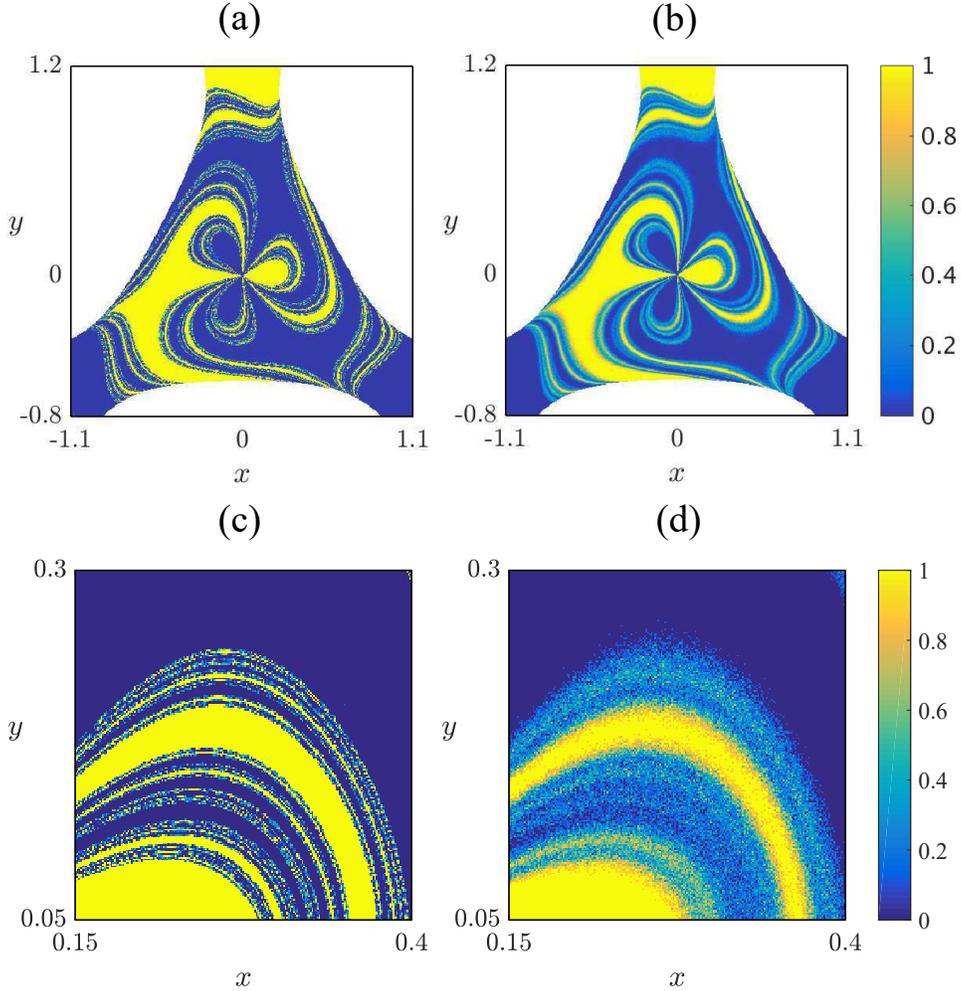}
	\caption{Color map showing the probability basin for the Exit 1 of the Hénon-Heiles system with energy $E=0.25$, and different noise intensities (a) $\xi=10^{-10}$ and (b) $\xi=10^{-6}$. Panels (c) and (d) are magnifications of the probability basin of panels (a) and (b), respectively. The yellow color indicates maximum probability of finding the Exit 1, while dark blue implies zero probability. Intermediate colors refer to intermediate probabilities, as shown in the color bar. To generate these figures, we have computed $100$ times the exit of each initial condition in a $400\times400$ grid and we have represented the probability of the Exit 1.}
	\label{fig4}
\end{figure}

To illustrate the above, in Fig.~\ref{fig4} we use a color code map to represent the probability basin of the Exit 1 of the Hénon-Heiles system with energy $E=0.25$ and different values of the noise intensity. We 
do not show the probability basins of the Exits $2$ and $3$ because, due to the symmetry of the basin, they are a simply $2\pi/3$ rotation of the probability basin of the Exit 1. In panel $(a)$ and $(c)$ the almost negligible noise intensity $\xi=10^{-10}$ generates mainly extreme values in the probability, indicating an almost deterministic behavior. However, even a very small amount of noise of intensity $\xi=10^{-6}$ can generate an important source of uncertainty in the vicinity of the deterministic basin boundary. The highly predictable regions become uncertain and the 
probabilistic nature of the escapes manifests along an important part of the physical space. Hence, we can conclude that the effects of noise make 
even more difficult the prediction of the final state of the system.

\begin{figure}[htp]
	\centering
	\includegraphics[width=0.6\textwidth,clip]{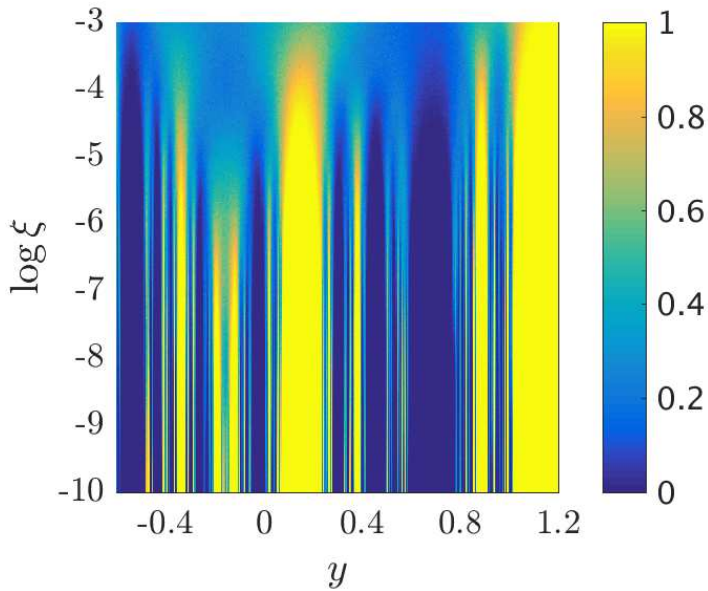}
	\caption{Evolution of the segment $x=0$ of the probability basin of the Exit 1 of the Hénon-Heiles system with increasing noise. The energy of the system is $E=0.25$, while the noise intensity is varied from $\xi=10^{-10}$ to $\xi=10^{-3}$. The color code is as described in the caption of Fig.~\ref{fig4}. The base of the log is $10$.}
	\label{fig5}
\end{figure}

As we increase the noise intensity, the uncertainty and the blurring of the probability basins also increase. To illustrate this, we represent in Fig.~\ref{fig5} the evolution of the segment $x=0$ of the probability basin of the Exit 1 when the noise intensity varies from $10^{-10}$ to $10^{-3}$. We have not considered higher noise intensities because they imply the dominance of the stochastic behavior and the trajectories are simply random walks. In the previous figure, we can observe how the regions where particles surely escape through Exit 1 decrease in size as noise increases. For $\xi=10^{-3}$ the only region where this happens is the one near to $y=0$, which is formed by initial conditions above the Lyapunov 
orbits \cite{Contopoulos90} that do not take place in the chaotic scattering process.

\section{Uncertainty basin}\label{BU}

\begin{figure}[htp]
	\centering
	\includegraphics[width=0.8\textwidth,clip]{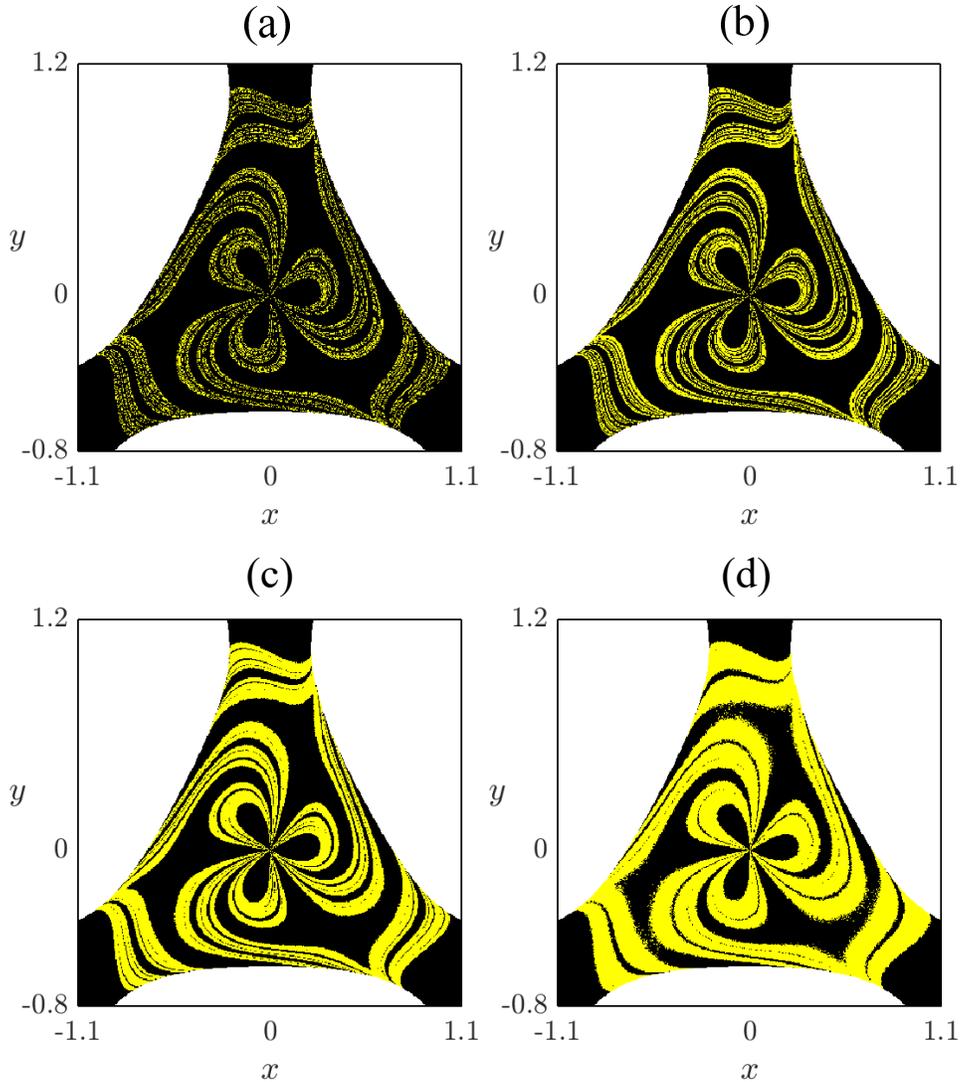}
	\caption{Uncertainty basins of the Hénon-Heiles system with energy $E=0.25$, and different noise intensities (a) $\xi=10^{-5}$, (b) $\xi=10^{-6}$, (c) $\xi=10^{-8}$ and (d) $\xi=10^{-10}$. To generate these figures we have computed the exit of each initial condition $100$ times and labeled the initial condition as certain (black) or uncertain (yellow).}
	\label{fig6}
\end{figure}

The probability basin representation is a useful tool to visualize the probability distribution of the exits along the phase space. Nevertheless, in many contexts we are not specifically interested in the probability, but on the unpredictability of the escape. For this reason, in deterministic chaotic scattering the boundary of the exit basins plays a central role when analyzing and quantifying the unpredictability of a system. It was 
not in vain that a large number of tools and concepts have been developed 
just to understand the basin boundary and its sometimes rich fractal structure. However, in the case of noisy chaotic scattering the basin boundary is not well defined since the set of points that define it changes for different simulations. For this reason it is necessary to change again our point of view and understand the uncertain initial conditions as the set of initial conditions that can change for different launchings. We call 
this set \textit{uncertainty basin}, $U_B$. We recall again that no uncertainty in the initial condition is considered here and the source of uncertainty is all due to the influence of the noise on the chaotic dynamics. 
In order to construct the uncertainty basin, we follow a similar method as in the case of the probability basins. First, we define a grid of $P$ initial conditions in all the phase space and we launch each of them $N$ times, labeling an initial condition as uncertain if the exit through which it escapes changes at least in one simulation. To illustrate the result, we simply associate a different color to the certain and the uncertain initial conditions. In Fig.~\ref{fig6} we show in yellow the uncertainty basin of the Hénon-Heiles with energy $E=0.25$ and different noise 
intensities (a) $\xi=10^{-10}$, (b) $\xi=10^{-8}$, (c) $\xi=10^{-6}$ and (d) $\xi=10^{-5}$. It can be observed at naked eye that the uncertainty basin is located in the vicinity of the stable manifold of the chaotic saddle ($W_s$) (i.e. the boundary of the exit basins) and grows with 
increasing noise. This means that some initial conditions that were predictable in the deterministic system due to the large distance to the basin 
boundary become uncertain due to the effects of noise. If we consider an initial condition sufficiently close to the stable manifold of the chaotic saddle, a small perturbation can move it from one basin to another, so it is intuitive that the uncertainty basin lies close to $W_s$. Moreover, 
we can establish that

\begin{equation}
	\lim_{\xi\to 0} U_B = W_s,
\end{equation}
which constitutes an accurate method to determine the structure of $W_s$.

The uncertainty basin is a faithful portrait of the unpredictability of the system in the presence of a particular noise intensity. By simply computing and exploring the uncertainty basin, we can conclude which regions are uncertain and which ones remain predictable under the effects of noise.

\begin{figure}[htp]
	\centering
	\includegraphics[width=1\textwidth,clip]{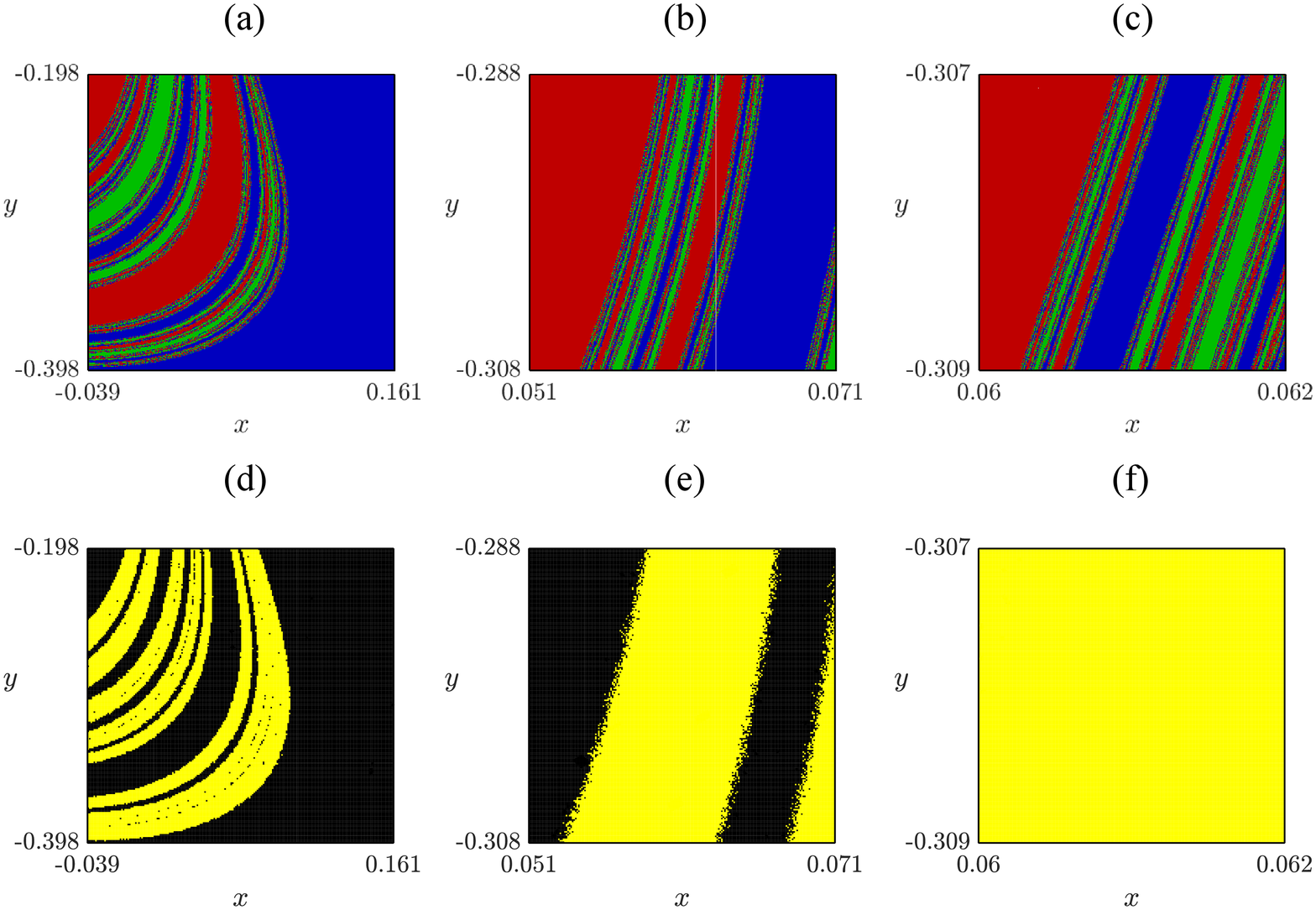}
	
	\caption{(a,b,c) Magnifications in the exit basins of the Hénon-Heiles in the physical space with energy $E=0.25$ (See Fig.~\ref{fig1}) and (d,e,f) same regions but in the uncertainty basin with noise intensity $\xi = 10^{-6}$ (See Fig.~\ref{fig6}(c)). It is clear that the uncertainty basin grows from the boundary of the exit basin, connecting different regions and generating a smooth geometry.  }
	\label{fig7}
\end{figure}

Even if the uncertainty basin is a set that emerges from the fine structure of the stable manifold of the chaotic saddle, as we increase the noise 
its once rich and complex structure becomes blurred, with the consequently lost of fractality. All those ``threads" that make up the boundary widen 
in the uncertainty basin and end up coming together. As a consequence, they define extensive regions of smooth geometry. This result is shown in Fig.~\ref{fig7}, where several magnifications on the exit basin and the uncertainty basin are represented. Meanwhile, the geometry of the basin boundary of the exit basin is not simplified when magnified. As a matter of fact, the smooth nature of the uncertainty basin manifest on a finite scale. One of the most significant implications of the absence of fractality in the uncertainty basin is that its size is independent of its resolution, which allows us to establish an absolute fraction of uncertain initial conditions.

\section{Quantifying the final state sensitivity} \label{QFSS}

In the previous sections we have introduced the concepts of probability basin and uncertainty basin, in order to give a graphic and qualitative understanding of the unpredictability on noisy chaotic scattering. However, 
in many physical problems we need a quantitative measure of the unpredictability in order to analyze the effect of a parameter or to compare different systems. In particular, even if two systems have the same fraction of uncertain initial conditions for a particular value of the noise, we do 
not have any guarantee that both have the same sensitivity to the noise, i.e., the fraction of uncertain initial conditions will increase in the same amount when the noise intensity increases. From the perspective of the prediction of the system, it is preferable that the uncertainty grows slowly with increasing noise. In this section, we develop a method to quantify the final state sensitivity to noise.

In deterministic systems one of the most common and useful measures is the uncertainty exponent, that can be obtained by using the uncertainty algorithm \cite{Grebogi83}. The method consists of launching several initial conditions under some uncertainty $\delta$. Then, we say that 
an initial condition is uncertain under an uncertainty $\delta$ if the asymptotic behavior is affected by the error. The fraction of uncertain initial conditions is then related to $\delta$ through a power law
\begin{equation}
	f_u(\delta) \sim \delta^{\alpha}, \label{ue}
\end{equation}
where $\alpha$ is the uncertainty exponent.

In noisy chaotic scattering system we have two sources of uncertainty: the error in the initial condition and the noise. For this reason, the fraction of uncertain initial conditions, $f(\delta,\xi)$, does not satisfy $f(0,\xi)=0$ since some initial conditions are uncertain even without considering an error. These initial conditions are those that change the exit for different launchings. These initial conditions will be uncertain $\forall \delta$, so they do not increase or decrease when changing $\delta$. This makes Eq.~(\ref{ue}) invalid in noisy systems. In fact, since the uncertainty exponent is defined in the limit $\delta \to 0$, the fraction of uncertain initial conditions for a fixed and not negligible noise intensity will always be $f(\delta,\xi)=f(\xi)=k$, where $k$ is a constant. This implies that $\alpha=0$ in noisy chaotic scattering systems.

Here, we are interested in the effects of noise, so we consider $\delta=0$ and we set that an initial condition is uncertain if it can escape through different exits for different launchings. Under this assumption the fraction of uncertain initial conditions obeys the law

\begin{equation}
	f_u(\xi) \sim \xi^{\beta}, \label{uen2}
\end{equation}
where $\beta>0$ is a magnitude that characterizes the intrinsic uncertainty of the noisy system, and we call it \textit{noise-sensitivity exponent}.

As $\beta$ approaches to zero a reduction in the noise intensity has less 
effect in decreasing the fraction of uncertain initial conditions, so we say that the system has high sensitivity to noise. In the limit $\beta=0$ a reduction in the noise intensity does not decrease the fraction of uncertain initial conditions at all. On the other hand, for high values of 
$\beta$ the fraction of uncertain initial conditions tends to zero for low noise intensities and very large noise intensities are needed in order to increase the unpredictability. Hence, we say that the system has low sensitivity to the noise. We have computed the parameter $\beta$ for several open Hamiltonian systems and their result ranges from $0.05$ to $0.5$. 
Furthermore, we have observed that extreme values appear only where the basin boundary occupies almost all the phase space ($\beta<0.05$) or in basins that have a unique smooth boundary ($\beta>0.5$).

In order to obtain $\beta$, we begin computing $f_u(\xi)$ for different values of $\xi$. In particular, for a fixed value of the noise intensity we launch $P$ initial conditions $N$ times, and we compute the fraction of 
uncertain initial conditions. We repeat this procedure for several values 
of $\xi$ and we proceed to a $\log-\log$ representation that will satisfy

\begin{equation}
	\log_{10}{f_u(\xi)} =\beta\log_{10}{\xi} + \mbox{C}, \label{uen2}
\end{equation}
where $C$ is a constant and the slope of the straight line is $\beta$.

\begin{figure}[htp]
	\centering
	\includegraphics[width=0.6\textwidth,clip]{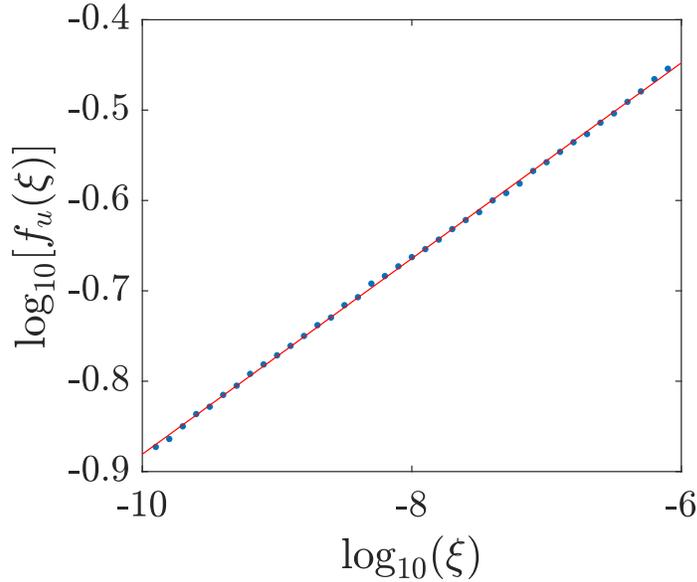}
	\caption{Logarithm of the fraction of uncertain initial conditions in function of the logarithm of the noise intensity of the Hénon-Heiles system with energy $E=0.25$. The parameter $\beta$ is estimated to be $\beta=0.109$. To generate this figure, we have used $40$ different values of the noise intensity. For every noise intensity the fraction of uncertain initial conditions has been computed after launching $100000$ initial conditions ($100$ times each one)}
	\label{fig8}
\end{figure}

The quality of the linear fit turns out very well for all the simulations 
that we have carried out (linear correlation coefficient $r>0.999$). As an example, we show the resulting linear fit for the case of the Hénon-Heiles system with energy $E=0.25$ in Fig.~\ref{fig8}. The noise-sensitivity exponent is estimated to be $\beta=0.109$. If we increase the energy to $E=0.45$ the result is $\beta=0.2309$, so that the final state sensitivity due to the noise is higher for $E=0.25$. Certainly, this 
case can be easily understood due to the big difference in the size of the basin boundaries (see Fig.~\ref{fig1}).

Another interesting application of the noise-sensitivity exponent is to compare the sensitivity to the noise for different systems. To illustrate this, we have chosen the four-hill system with energies $E=0.01$ and $E=0.1$. The exit basins of the conservative system for these values of the energy have been shown in Fig.~\ref{fig2}. In these cases the noise-sensitivity exponent is estimated to be $\beta=0.078$ ($E=0.01$) and $\beta=0.2902$ ($E=0.1$). Hence, for the lower energy value the system is more sensitive to the noise than the Hénon-Heiles for $E=0.25$, while for the higher energy value the four-hill system is less sensitive than the Hénon-Heiles with $E=0.45$.

\section{Conclusions }\label{Conclusions}

In summary, our research reveals that the noise, even if the intensity is 
very weak, has important implications on the predictability of chaotic scattering systems. We have shown that in order to have a deeper understanding of the effects of noise, some concepts and tools of nonlinear dynamics and chaos should be revisited. In particular, the usual ways to represent the exit basins and to compute the uncertainty exponent are meaningless in the presence of noise. Throughout this work we have adopted a probabilistic point of view and developed the concepts and methods to compute the probability basin, uncertainty basin and noise-sensitivity exponent. Under all these concepts underlies the idea that an initial condition is uncertain if it can change its exit for different launchings. The probability and uncertainty basins allow a qualitative analysis of the probability distribution of the escapes and the structure of the uncertain regions, 
respectively. On the other hand, the noise-sensitivity exponent gives a measure to quantify the final state sensitivity in presence of noise. This 
tool can be useful to analyze how the sensitivity to the noise  manifests 
itself for different dynamical systems, as well as to see its behavior when a parameter of the system varies. 

We expect that the results shown in this work could be helpful in providing new concepts and tools to future research concerning chaotic scattering problems.

\section*{Acknowledgments}
We dedicate this paper to the memory of Prof. Tito Arecchi with whom we had the pleasure of collaborating in different nonlinear dynamics projects. This work has been financially supported by the Spanish State Research Agency (AEI) and the European Regional Development Fund (ERDF) under Project No. PID2019-105554GB-I00.

\end{document}